# ITERATIVE LEARNING CONTROL – DEEP DIVE

S. R. Koscielniak*, TRIUMF, Vancouver, B.C., Canada

*Abstract*

The stability and convergence of an Iterative Learning Controller[1] (ILC) may be assessed either by directly iterating the equations for a variety of inputs, or by finding the eigenvalues ($\lambda$) of the iterated system, or by forming the Z-transform and applying pole-zero or equivalent root locus. Two often-used criteria are (i) Asymptotic Convergence (AC) of the difference vectors, and (ii) monotonic convergence (MC) of the vector norm. The latter (MC) has a Z-domain counterpart. In this paper we apply all three methods and both convergence tests to a simple plant with an ILC wrapper. One, two and three-term learning functions are used. We can then ask the questions: do all the tests work, and do they agree on the stability?

## INTRODUCTION

Concerning this study, "*testing the tests*", the author wished to verify on a realistic & difficult (but tractable) problem that all analysis tools can be successfully applied to all parts of the problem. Good to check the theory against analytic cases before resorting only to numerical methods – which can only address specific initial conditions and parameters. Contrastingly, analysis can address all conceivable parameters – if the problem is tractable. Analysis is more powerful and complete than simulation.

Additional motivation: Are z-domain (Z-D) and eigenvalue domain ($\lambda$-D) convergence tests identical? Do they agree about the convergence domain estimated from iteration-domain (I-D) simulations?

### Model of the Plant

The "plant" [2] is an RF cavity (time constant $\tau_c = 1/a$) sandwiched by a zero-order hold (ZOH) sampling system (rate $\rho_s = 1/\tau_s$) with a PI (proportional gain Kp & integral gain Ki per second) controller. Let U be the product $U = a\cdot\tau_s = \tau_c\,\rho_s$. When pole-zero cancellation is used (as here) for the PI controller, then Ki=a.Kp. The plant is wrapped by an iterative learning control (ILC) with gain $\nu$. There are two cases: (i) proportional control alone, which is treated here; and (ii) full PI control. Results for the latter have been obtained, are not reported here due to lack of text space.

### Stability Analysis in Z-domain

Let z denote the argument of the Z-transform, the Laplace transform for discrete, sampled systems. System is stable if the poles of the closed loop transfer function P(z) are inside the unit circle. Equivalent: system stable if the zeros of the denominator of P(z) are inside the unit circle. This may be determined either by examining zeros or plotting the root locus, whichever is easier.

Let $C[z] = K_p + \frac{K_i \tau_s z}{(z-1)}$ and $G(z) = \frac{(e^U - 1)}{(ze^U - 1)}$

The closed loop transfer function is: $P(z) = \frac{G(z)}{1 + C(z)G(z)}$.

All the following deals with the single pole problem when Ki=0. The analysis [3] is simplified if we transform from variables U and Kp to A and B, as follows:

$A = 1 - e^{-U}$ and $B = e^{-U}(1 + K_p) - K_p$.

The ranges of A and B are $0 < A < 1$ and $-1 < B < 1$. If $U \ll 1$ and Kp>0, B is close to unity and A close to zero. When Ki=0, the closed loop gain function becomes

$P(z) = \frac{-1 + e^U}{-1 - K_p + e^U K_p + e^U z}$ which may be written $\frac{A}{z - B}$.

The pole z has to lie within |z|<1. If there were no other constraints, the conditions {B >-1, B >0, B <1} imply, respectively, that

$\{K_p < \frac{1 + e^U}{-1 + e^U}, K_p < \frac{1}{-1 + e^U}, K_p \geq -1 \}$.

The first and second conditions have different implications in time domain. The second condition supersedes the first.

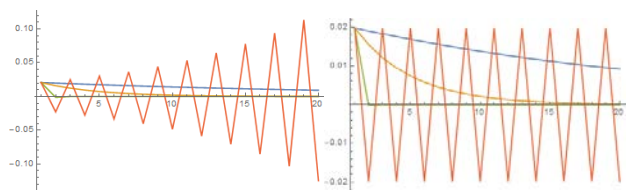

Fig.0: Behaviour within a trial. ILC off. Left: First gain condition exceeded. Right: Second condition exceeded.

Evidently, large gain Kp can be achieved if U is small; which occurs when the sampling rate is high and/or the cavity time constant is small.

### Stability Analysis of ILC in Z-domain

Transforms take a complicated function in one space into a simpler function in the transform space. The Fourier, Laplace and Z domains all share the advantage that one does not have to worry about time ordering: operators commute and convolutions become products. Stability tests in these domains do not rely upon the assumption of geometric series, and (in principle) are more general than the tests employed in $\lambda$-domain. Let $\nu$ be the learning gain.

In the extended Z-domain of trials and iterations, the closed-loop iteration equation becomes $u_{j+1}(z) = T(z)u_j(z)$. Not with $T(z) = Q(z)[1 - \nu L(z)P(z)]$, but with $T(z) = Q(z)[1 - \nu L(z)zP(z)]$. The additional power of z accounts for the lift that is applied to the matrix representation of P to compensate for the ZOH. Stability analysis in z-domain proceeds by substituting $z = \mathrm{Exp}[i\Theta]$ (the unit circle) into T, and then forming the locus of T in the complex plane as Θ varies from 0 to π. The condition for monotonic convergence of the norm of the vector (MC) which records the behaviour during a trial is Supremum[Abs[$T[e^{iΘ}]$]] < 1.

---

* shane@triumf.ca

Let R=|T(z)| and Tanφ = Im[T(Θ)]/Re[T(Θ)]. System is stable if T remains within the unit circle Exp[i φ] for all Θ. (i.e. R ≤ 1 ). If R=1, then dR/dφ must be identically zero; and $d^2R/dφ^2$ <0. [For comparison, on the unit circle (itself) all derivatives of R w.r.t. φ must be zero.] Because |T(φ)| <1 is (generally) not a unit circle, so it follows that when R(φ)=1 odd derivatives w.r.t. φ must be zero and even derivatives < 0. Typically T(Θ) is less complicated than T(φ). Fortunately, we can often work with T(Θ) because:
dR/dφ = (dR/dΘ)(dΘ/dφ) = 0 if dR/dΘ=0 or dΘ/dφ=0;
$d^2R/dφ^2$ = $(d^2R/dΘ^2)(dΘ/dφ)^2$ + $(d^2Θ/dφ^2)(dR/dΘ)$ < 0 if dR/dΘ=0 and $(d^2R/dΘ^2)$ < 0.

*Asymptotic Convergence*

Causal Learning Functions (Look Back)
The matrix operator representation of P for causal learning functions are Toeplitz, and their eigenvalues are equal to the elements on the diagonal, which are identical. In this particular case, the condition for asymptotic convergence (AC) is λ=1-A<1 independent of B.

Noncausal Learning Functions (Look Ahead)
Contrastingly, the eigenvalues of the iteration matrix λ(**A**) for look-ahead type of **L** are not degenerate and are spread. The condition for asymptotic convergence is all |λ(**A**)|<1.

*1-term learning L=vI & Ki=0*

L(z)=v: use only information from element *k* of the previous trial vector when updating element *k* at the start of the next trial. $T = 1 - \frac{Ave^{iθ}}{-B+e^{iθ}}$ . T and R=|T| is largest at phases θ=0,π. (θ=0 →DC; θ=π → Nyquist frequency = $ρ_s$/2.) Imposing the condition dR/dΘ =0 at those phases leads to $(-1 < B ≤ 0 \ \& \ 0 < Av < 2 + 2B)$ or
$$(0 < B < 1 \ \& \ 0 < Av < 2 - 2B)$$
And hence $0 < Av < 2 \ \& \ -1 + \frac{1}{2}Av < B < 1 - \frac{1}{2}Av$
And hence $0 < v < 2(1 + Kp)$ and $-1 < Kp < \frac{1}{-1+e^U}$
The latter is the gain limit with no ILC. These Z-D conditions are confirmed by I-D and λ-D Figs.[1-3], AC Fig.4

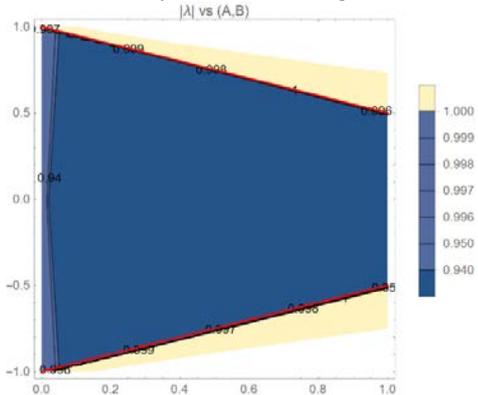
Figure 1: Eigenvalues λ($A^TA$) vs (A,B) for L=I (i.e. v=1.) MC-domain |λ|<1 shown blue.

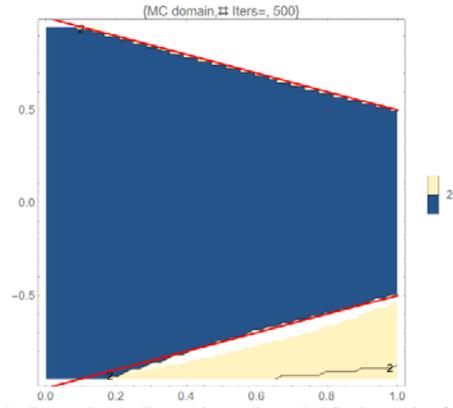
Figure 2: Based on direct iteration: MC-domain for **L=I**.

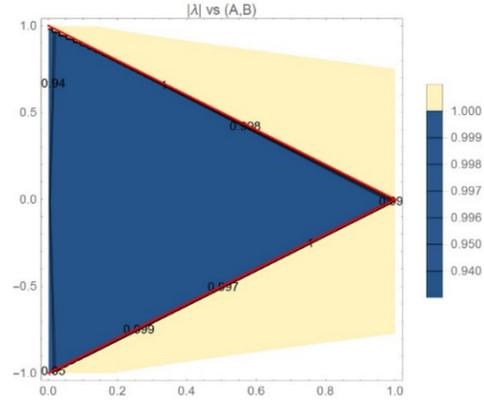
Figure 3: Eigenvalues λ($A^TA$) vs (A,B) for **L=vI** & v=2. Confirms the dependence on learning gain v.

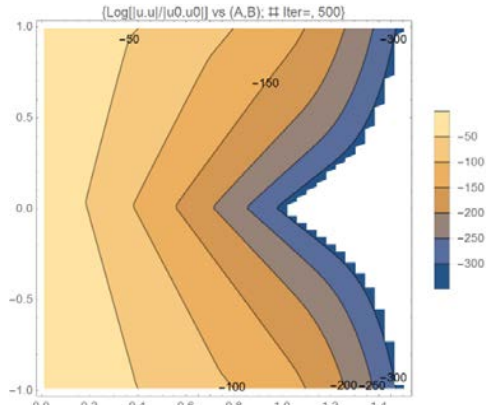
Figure 4: Based on direct iteration: AC-domain for **L=I**. Basically it is the entire area |A|<1 & |B|<1. Agrees with expectation of λ=1-A independent of B.

*2-Term L, Look-Back 1 Step; & Ki=0*

L(z) = v(1+1/z) is causal: when updating each element of the super-vector, use only information from present and previous time step. $T = 1 - \frac{Ave^{iθ}(1+e^{-iθ})}{-B+e^{iθ}}$
T and R=|T| is largest at phases θ=0,π; imposing dR/dΘ =0 at those phases, leads to conditions:
$$0 < Av < 2 \ \& \ -1 < B < 1 - Av$$
We add the constraint B>0, leading to $-1 < Kp < \frac{1}{-1+e^U}$ and $v < (1 + Kp)$ . The first repeats the stability without ILC, the second limits the learning gain. These Z-D conditions are confirmed by I-D and λ-D, Figs. [5-7] & AC Fig.8

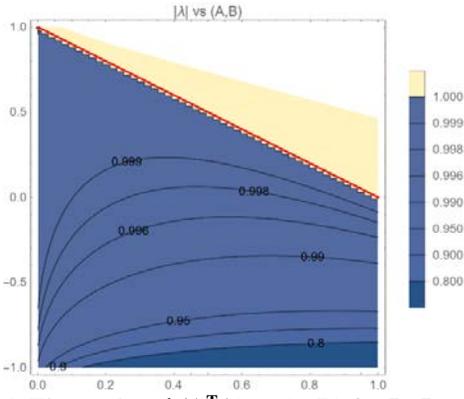
Figure 5: Eigenvalues λ($A^T A$) vs (A,B) for **L=I+↓** (i.e. ν=1.) MC-domain |λ|<1 shown blue.

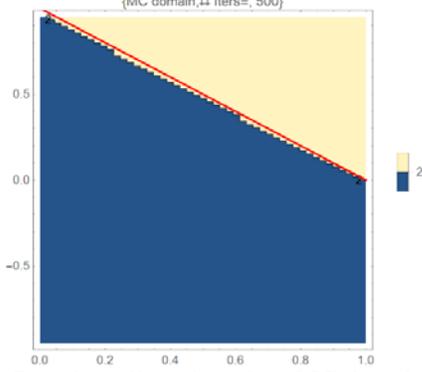
Figure 6: Based on direct iteration, MC-domain vs (A,B) for **L=I+↓** (i.e. ν=1.)

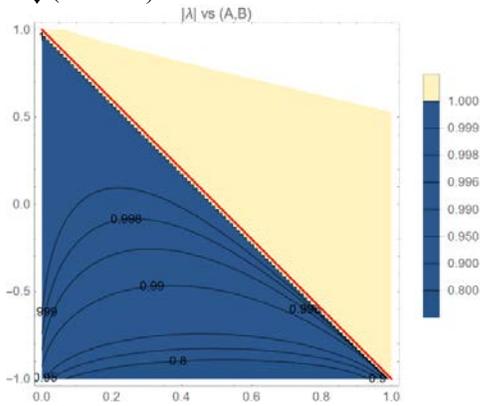
Figure 7: Eigenvalues λ($A^T A$) versus (A,B) for **L=2(I+↓)** (i.e. ν=2.) Confirms the dependence on learning gain ν.

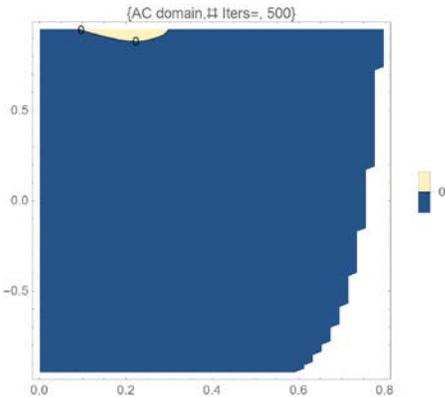
Figure 8: Based on direct iteration, estimate of AC-domain versus (A,B) for **L=I+↓** (i.e. ν=1.).

Fig.8: Yellow area in top left corner has not converged after 500 iterations. However, AC domain will eventually become the entire area |A|<1 & |B|<1 – but extremely slowly.

*2-Term L, Look-Ahead 1 Step; & Ki=0*

L(z) = ν(1+z). Element *k* of the next trial is updated using information from elements *k* and *k-1* of previous trial.

$$T = 1 - \frac{A\nu e^{i\theta}(1+e^{i\theta})}{-B+e^{i\theta}}$$

T is large in four directions, including the directions θ=0,π. We impose the condition dR/dΘ =0 at those phases, and minimize with respect to Aν. After some work, we find the sufficient condition:

$$0 < A\nu < 1 \,\&\, (-1+A\nu)/3 \le B \le 1 - A\nu$$

We add the constraint B>0, leading to

$$-1 < Kp < \frac{1}{-1+e^U} \text{ and } \nu \le (1+Kp)$$

These Z-D conditions are confirmed by I-D and λ-D Figs. [9-10] and AC Figs. [11 &12].

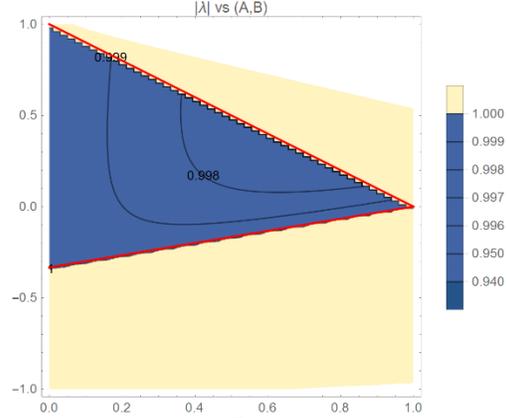
Figure 9: Eigenvalues λ($A^T A$) versus (A,B) for **L=(I+↑)** (i.e. ν=1). MC-domain |λ|<1 shown blue.

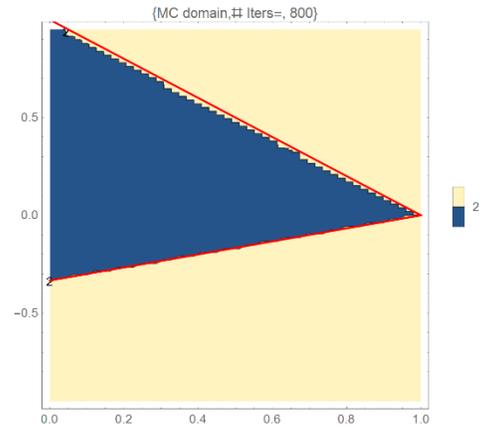
Figure 10: Based on direct iteration, MC-domain versus (A,B) for **L=I+↑** (i.e. ν=1.)

Testing for Monotonic Converge in Iteration Domain
For a particular point (A,B) one must record the iteration # on which MC started, and the iteration number when MC stopped. If latter is greater than former, then (A,B) is not MC. If former is first iteration, and MC never stops, then point (A,B) is MC. Note: before convergence starts, vector-norm of transients may each astronomically large values!

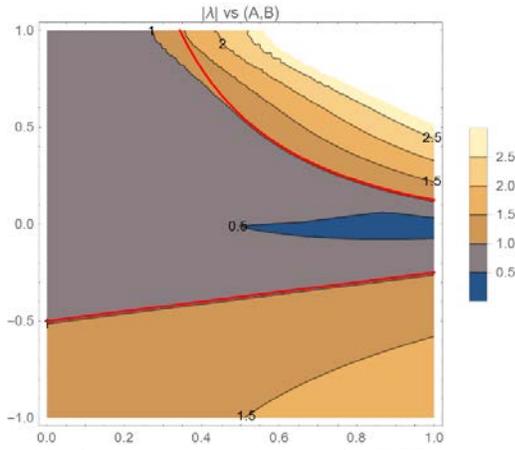

Figure 11: Based on eigenvalues of **A**=(**I-PL**), estimate of AC-domain versus (A,B) for **L=(I+↑)**.

Lower & upper limits $B > \frac{1}{2}(-1 + \frac{A}{2})$ & $B < \frac{(2-A)^2}{8A}$.

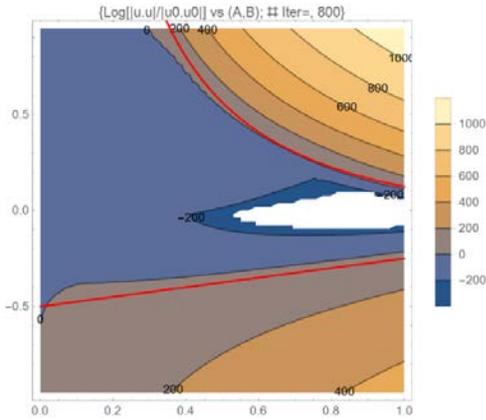

Figure 12: Based on 800 iterations, estimate of AC-domain versus (A,B) for **L=(I+↑)**. Log plot of ratio, therefore negative values mean "converged".

*3-Term Symmetric Learning*

L(z) = ν(1+z+1/z); use information from present, previous and next step within a trial $T = 1 - \frac{Ave^{i\theta}(1+e^{-i\theta}+e^{i\theta})}{-B+e^{i\theta}}$
T is large in four directions. We consider θ=0, π, π/2 leading to the contradiction: Aν>0 & Aν<0; and prediction: iterations are *not* monotonic convergent. In fact, they are not even asymptotically convergent (based on direct iteration). However, taking a learning function with unequal gains for I versus ↓ or ↑, such as **L=I + (1/2) (↑+↓)**, restores stability and convergence. These Z-D conditions are confirmed by I-D and λ-D, Figs. [13 & 14].

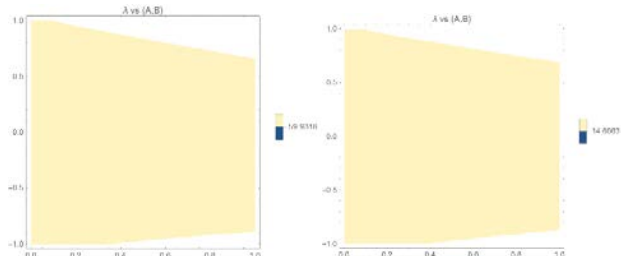

Figure 13: Eigenvalues λ(**A^T A**) versus (A,B) for **L=(I+↓+↑)**. Left is case ν=1. Right is case ν=2.

Fig.13-left: Yellow =λ>1.01, white=λ>60. No eigenvalues |λ|<1; therefore no MC domain.
Fig.13-right: Yellow =λ>1.01, white= λ>14.6. No MC domain. Confirms dependence on iteration gain ν.

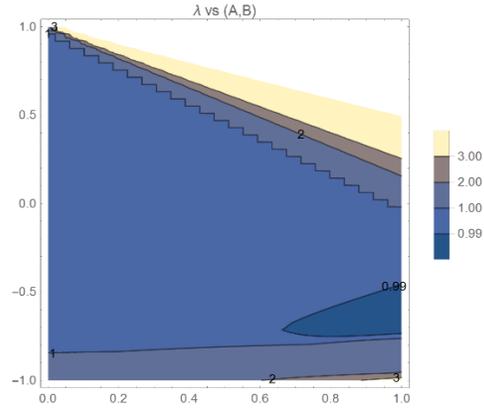

Figure 14: Eigenvalues λ(**A^T A**) versus (A,B) for **L=I+(1/2)(↓+↑)**. MC-domain |λ|<1 shown blue.

*3-Term Look Ahead*

L=ν(1+z+z²). T is large at six phases. R=1 when θ= 2π/3. We consider θ=0,π, π/2 leading to necessary, but possibly insufficient, conditions:

$$B < 1 - \frac{3A\nu}{2} \text{ and } B > \frac{1}{2}(1 + \frac{A\nu}{2})$$

These Z-D conditions are confirmed by I-D and λ-D, Figs. [15 &16]; and AC conditions Figs. [17 & 18].

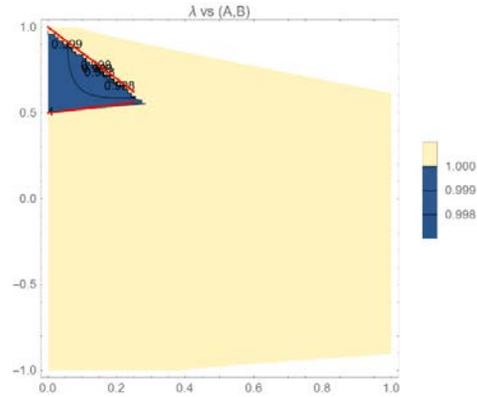

Figure 15: Eigenvalues λ(**A^T A**) versus (A,B) for **L= (I+↑+↑)** i.e. ν=1. MC-domain |λ|<1 shown blue.

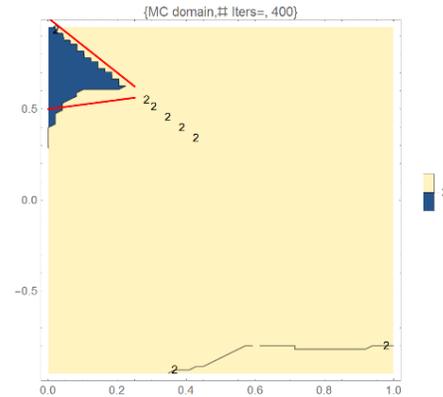

Figure 16: Based on direct iteration, MC-domain vs (A,B) for **L=I+↑+↑** (i.e. ν=1.)

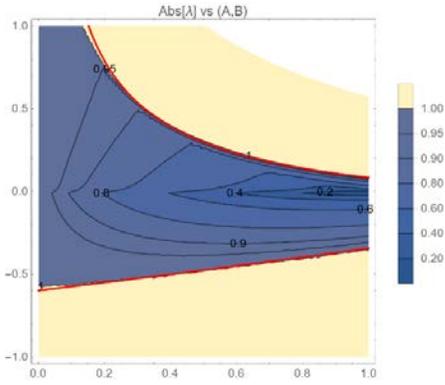

Figure 17: Eigenvalues of λ(**A**) versus (A,B) for L=(I+↑+↑) AC domain shown in blue.

From the eigenvalues of **A**, the domain of asymptotic converge is approx $B > -0.6(1 - \frac{A}{2})^{0.8}$ and $< \frac{(2-A)^2}{12A^{2/3}}$.

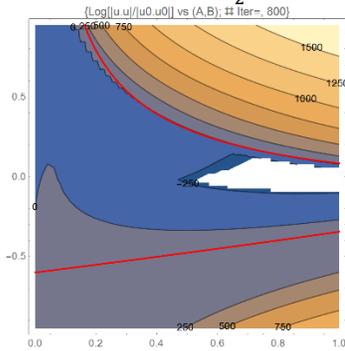

Figure 18: Based on 800 iterations, estimate of AC-domain versus (A,B) for **L=(I+↑+↑)**. Log plot, therefore negative values mean "converged".

*3-Term Look Back*

L = ν(1+ 1/z+ 1/z² ); use information from present and two previous time steps.

$$T = 1 - \frac{Ave^{i\theta}(1 + e^{-i\theta} + e^{-2i\theta})}{-B + e^{i\theta}}$$

T and R=|T| is large at four phases, including θ=0,π. R=1 when θ= 2π/3. We consider θ=0, π, π/2 leading to necessary, but possibly insufficient, conditions

$B < \frac{1}{42}(-16 - 7A)$ and $B > \frac{1}{42}(-28 + 7A)$

These Z-D conditions are confirmed by I-D and λ-D, Figs. [19 &20]; and the causal AC condition Fig. [21].

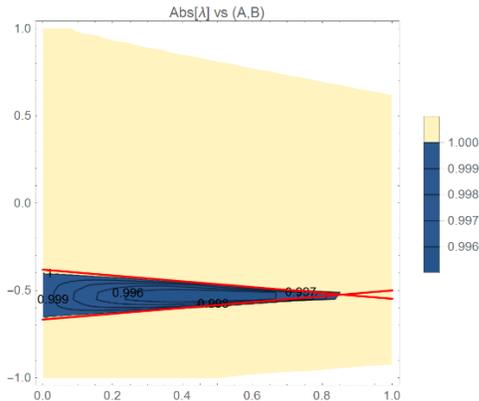

Figure 19: Eigenvalues λ(**A**$^T$**A**) versus (A,B) for **L= (I+↓+↓)** i.e. ν=1. MC-domain |λ|<1 shown blue.

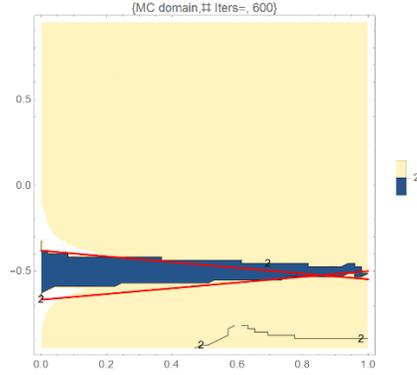

Figure 20: Based on direct iteration, MC-domain vs (A,B) for **L=I+↓+↓** (i.e. ν=1.) MC domain shown blue.

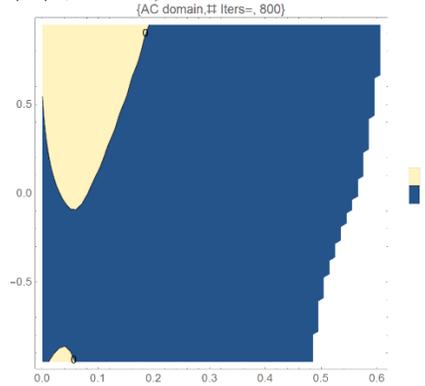

Figure 21: Based on 800 iterations, estimate of AC-domain versus (A,B) for **L=(I+↓+↓)**. Blue and white areas mean "converged". The blue area expands as the number of iterations is increased – but <u>extremely</u> slowly.

Why is Z-domain stability test a condition for MC?

The Z (or Laplace) domain tells us about stability within the trial interval including the internal dynamics introduced by the learned time function. Because it is about stability within the trial it cannot say anything about iteration eigenvals. However, it can tell us if system settles down, and if so then the iterations must have converged. The time domain analog of |x[n]|² is integral[f[t]^2,{t,nT,(n+1)T}]. The connection between time domain and Fourier domain integral of modulus is Plancherel's theorem. Hence the Z-domain test will be about convergence of |x[n]|².

## CONCLUSION

Author has successfully applied Z-domain and time-domain (t) analysis to verify the stability of the system <u>without</u> ILC. Author has applied Z-domain and eigenvalue-domain (λ) and Iteration-domain analysis to study the convergence/stability of the system <u>with</u> ILC.

Author has found: Z-domain MC test gives results (almost) identical to λ-domain MC test; and both are confirmed by direct iteration. λ-domain AC test typically very different from the λ-domain MC test (as of course, is to be expected). For parameters which pass the λ-domain AC test but fail the λ-domain MC test, iteration-domain tests typically show there are "learning transients", which may occur either for "look ahead" or "look back" learning.